\documentclass[showpacs,amsmath,amssymb,preprint]{revtex4}
\usepackage{epsfig,dcolumn,bm}

\begin{document}

\title{$NK\pi$ molecular state with $J^{\pi}=\frac{3}{2}^-$ and $I=1$}

\author{F. Huang$^{1,2,3}$}
\author{Z.Y. Zhang$^2$}
\author{Y.W. Yu$^2$}
\affiliation{\small
$^1$CCAST (World Laboratory), P.O. Box 8730, Beijing 100080, PR China \\
$^2$Institute of High Energy Physics, P.O. Box 918-4, Beijing 100049, PR China\footnote{Mailing address.} \\
$^3$Graduate School of the Chinese Academy of Sciences, Beijing,
PR China }

\begin{abstract}
The structure of the moleculelike state of $NK\pi$ with
spin-parity $J^{\pi}=\frac{3}{2}^-$ and isospin $I=1$ is studied
within the chiral SU(3) quark model. First we calculate the $NK$,
$N\pi$, and $K\pi$ phase shifts in the framework of the resonating
group method (RGM), and a qualitative agreement with the
experimental data is obtained. Then we perform a rough estimation
for the energy of $(NK\pi)_{J^{\pi}=\frac{3}{2}^-,I=1}$, and the
effect of the mixing to the configuration $(\Delta
K)_{J^{\pi}=\frac{3}{2}^-,I=1}$ is also considered. The calculated
energy is very close to the threshold of the $NK\pi$ system. A
detailed investigation is worth doing in the further study.
\end{abstract}


\maketitle

For the past few years, many low-energy baryonic resonances have
been explained as baryon-meson molecule resonance states within the
chiral unitary approach \cite{kaiser95}. To understand this kind of
mechanism on the quark level and to compare further the similarities
and the differences between the results obtained on the hadron level
and those on the quark level is very significant. Recently we have
extended our chiral SU(3) quark model from the study of
baryon-baryon scattering processes to the baryon-meson systems by
solving the resonating group method (RGM) equation
\cite{huangf04prc1,huangf04prc2,huangf05prc1}. We found that some
results are similar to those given by the chiral unitary approach
study, such as that both the $\Delta K$ system with isospin $I=1$
and the $\Sigma K$ system with $I=\frac{1}{2}$ have quite strong
attractions. We also studied the phase shifts of $\pi N$ and $\pi
K$, and got reasonable fit with the experiments in the low energy
region \cite{huangf05ctp}. Because our calculation is on the quark
level, the parameters can be fixed by the baryon masses and the $NN$
(or $KN$) scattering, and thus the free parameters are largely
reduced. These encouraged us to investigate more baryon-meson
systems by use of the same group of parameters.

The most interesting case is the system with strangeness
${\cal{S}}=+1$, this is because for this five-quark system there is
no annihilation to both gluons and vacuum. At the same time, since
2003, several experimental collaborations reported that they
observed a new resonance $\Theta^+$, with positive charge and
strangeness quantum number ${\cal{S}}=+1$ \cite{hicks05}. The mass
of this particle is around 1540 MeV and its width is very narrow,
$\Gamma_{\Theta}<25$ MeV. But recently the situation has become more
complicated. Several high statistical $\gamma p$ experiments showed
their negative results \cite{woods05}, and, conversely, the
RHIC-STAR group reported the observation of a resonance state with
$+2$ charges and strangeness ${\cal{S}}=+1$, named $\Theta^{++}$,
and its mass is around 1530 MeV \cite{rhic05}. Although the
experimental result is unclear, the theoretical investigation of the
systems with strangeness ${\cal{S}}=+1$ from different mechanism to
understand if it is possible to have resonance states is still
alluring.

In Refs. \cite{huangf04prc1,huangf04prc2}, we studied the $KN$
scattering phase shifts in the chiral SU(3) quark model and further
in the extended chiral SU(3) quark model where the vector chiral
fields are included by performing a RGM calculation, and we obtained
considerably good agreement with the experimental data for the $S$,
$P$, $D$, and $F$ partial waves. From these works, we found that
there is neither $S$ nor $P$ wave resonance in $KN$ scattering for
both isospin $I=0$ and $I=1$ cases. This means that the dynamical
calculation in the chiral quark model can explain the $KN$ phase
shifts, but cannot obtain a resonance state.

We also noticed that though the interactions of $KN$ $S$ wave with
$I=0$ and $I=1$ are repulsive, the phase shifts of $N\pi$ $P$ wave
with $I=\frac{3}{2}$ and of $K\pi$ $S$ and $P$ waves with
$I=\frac{1}{2}$ show their interactions are attractive. This means
that probably the pion can be a medium to glue a kaon and a
nucleon together for some states with suitable quantum numbers.

In Ref. \cite{bicudo04}, the possibility of the $\Theta$ particle to
be a $NK\pi$ state with $J^\pi=1/2^+$ and $I=0$ has been studied.
But we notice that the $N\pi$ $P$-wave state with isospin
$\frac{3}{2}$ and total angular momentum $\frac{3}{2}$ has a strong
attraction, and the isospin $\frac{1}{2}$ $K\pi$ $S$- and $P$-wave
scattering phase shifts also show that their interactions are
attractive. These encourage us to study the possibility that there
will be a $NK\pi$ bound state near its threshold with special
quantum numbers because of the comparatively strong attractive
interaction of $N\pi$ and $K\pi$. In this work, we suggest that the
state of $NK\pi$ with $J^{\pi}=\frac{3}{2}^-$ and $I=1$ could be a
very interesting state. We estimate the energy of the $NK\pi$
three-body system, where the interactions of $KN$, $K\pi$ and $N\pi$
are obtained from the chiral SU(3) quark model. Further the
configuration mixing between $(NK\pi)_{J^{\pi}=\frac{3}{2}^-,I=1}$
and $(\Delta K)_{J^{\pi}=\frac{3}{2}^-,I=1}$ is also considered.
Although the calculation is only a rough estimate, the result is
quite interesting. It shows that when the configuration mixing is
included, the energy of the system is quite near the threshold of
$NK\pi$ (1572 MeV). This gives us a hint that the system of
$(NK\pi)_{J^{\pi}=\frac{3}{2}^-,I=1}$ could be a moleculelike bound
state. We also notice that if the system of
$(NK\pi)_{J^{\pi}=\frac{3}{2}^-,I=1}$ is really bound, its width
must be very narrow, because when and only when the pion of the
system is absorbed by the nucleon, can the decay process to a $K$
and a $N$ occur. All these features show that the
$(NK\pi)_{J^\pi={\frac{3}{2}}^-,I=1}$ state is an interesting state,
and the possibility of the $\Theta$ particle to be the
$(NK\pi)_{J^{\pi}=\frac{3}{2}^-,I=1}$ state cannot be excluded if
the $\Theta$ particle does really exist.

Because the state $(N\pi)_{l=1,s=\frac{1}{2},j=\frac{3}{2},
t=\frac{3}{2}}$ has a quite strong attraction, we thus start from
the coupling of $(N\pi)_{l=1,s=\frac{1}{2},j=\frac{3}{2},
t=\frac{3}{2}}$ with a kaon to construct the wave function of the
$NK\pi$ system. When the relative motion wave function between the
$(N\pi)_{l=1,s=\frac{1}{2},j=\frac{3}{2}, t=\frac{3}{2}}$ and the
kaon is taken to be $S$ wave, there are only two possible states:
isospin $I=1$ and $2$. A simple analysis shows that the state with
$I=1$ is more favorable. Therefore we suppose the $NK\pi$ system
with $J^{\pi}=\frac{3}{2}^-$ and isospin $I=1$ could be a molecular
state.

The wave function of the $NK\pi$ system with $J^{\pi}=\frac{3}{2}^-$
and $I=1$ can be written simply as follows:
\begin{eqnarray}
\Psi_{NK\pi}=[(N\pi)_{l=1,s=\frac{1}{2},j=\frac{3}{2},
t=\frac{3}{2}}K]_{L=1,S=\frac{1}{2},J=\frac{3}{2},I=1},
\end{eqnarray}
where $l$ is the orbital angular momentum of $N$ $\pi$ relative
motion, and $s$ and $t$ represent the spin and isospin of $N\pi$,
respectively. $L$ and $I$ are the total orbital angular momentum and
isospin of the $NK\pi$ system. When the orbital wave function is
taken to be the harmonic oscillator wave function, $\Psi_{NK\pi}$
can be expressed as follows:
\begin{eqnarray}
\Psi_{NK\pi}=\left\{\left(\sqrt{\frac{M_N}{M_{N\pi}}}\phi_{s}(\vec{r}_N)\phi_{p}(\vec{r}_{\pi})
-\sqrt{\frac{M_{\pi}}{M_{N\pi}}}\phi_{p}(\vec{r}_N)\phi_{s}(\vec{r}_{\pi})
\right)_{l=1,s=\frac{1}{2},
t=\frac{3}{2}}\phi_{s}(\vec{r}_K)\right\}_{J=\frac{3}{2},I=1}.
\end{eqnarray}
Performing a re-coupling expression of $\Psi_{NK\pi}$, one obtains
\begin{eqnarray}
\Psi_{NK\pi}=&&\left\{N\left[\sqrt{\frac{M_{\pi} M}{M_{N\pi}
M_{K\pi}}} \left(\sqrt{\frac{8}{9}}(K\pi)_{l=0,t=\frac{1}{2}}
+\sqrt{\frac{1}{9}}(K\pi)_{l=0,t=\frac{3}{2}}\right)\right.\right.\nonumber\\
&&+\left.\left.\sqrt{\frac{M_N M_K}{M_{N\pi} M_{K\pi}}}
\left(\sqrt{\frac{8}{9}}(K\pi)_{l=1,t=\frac{1}{2}}
+\sqrt{\frac{1}{9}}(K\pi)_{l=1,t=\frac{3}{2}}\right)\right]\right\}_{J=\frac{3}{2},I=1},
\end{eqnarray}
and
\begin{eqnarray}
\Psi_{NK\pi}=&&\left\{\pi\left[\sqrt{\frac{M_N M}{M_{N\pi}
M_{NK}}}\left(\sqrt{\frac{2}{3}}(NK)_{l=0,t=0}
+\sqrt{\frac{1}{3}}(NK)_{l=0,t=1}\right)\right.\right.\nonumber\\
&&+\left.\left.\sqrt{\frac{M_{\pi} M_K}{M_{N\pi} M_{NK}}}
\left(\sqrt{\frac{2}{3}}(NK)_{l=1,t=0}
+\sqrt{\frac{1}{3}}(NK)_{l=1,t=1}\right)\right]\right\}_{J=\frac{3}{2},I=1}.
\end{eqnarray}
Here
\begin{eqnarray}
M_{N\pi}&=&M_N+M_{\pi},~~~M_{K\pi}=M_K+M_{\pi},\nonumber \\
M_{NK}&=&M_N+M_K,~~~M=M_N+M_K+M_{\pi}.
\end{eqnarray}

Once the interactions of $NK$, $N\pi$ and $K\pi$ are obtained, the
energy of $(NK\pi)_{J^{\pi}=\frac{3}{2}^-,I=1}$ can be estimated by
using the above wave functions.

To get the interactions of $NK$, $N\pi$ and $K\pi$, we draw support
from the chiral SU(3) quark model that has been quite successful in
explaining the $NN$, $YN$ and $KN$ scattering data
\cite{zhangzy97,huangf04prc1,huangf04prc2}. The effective potentials
of $NK$, $N\pi$ and $K\pi$ can be extracted from the phase shift
fitting calculations. Here we briefly introduce the chiral SU(3)
quark model and show the phase shifts of $NK$, $N\pi$, and $K\pi$ in
the low energy region, and then we give a rough estimate for the
energy of the system $(NK\pi)_{J^{\pi}=\frac{3}{2}^-,I=1}$.

In the chiral SU(3) quark model, we introduce the coupling between
quarks and chiral fields to describe the low-momentum medium-range
nonperturbative quantum chromo-dynamics (QCD) effect. The
interacting Lagrangian ${\cal L}_{I}$ can be written as follows:
\begin{equation}
{\cal
L}_{I}=-g_{ch}\bar{\psi}(\sum\limits_{a=0}^{8}\sigma_{a}\lambda_{a}
+i\sum\limits_{a=0}^{8}\pi_{a}\lambda_{a}\gamma_{5})\psi.
\end{equation}
Here scalar nonet fields $\sigma_a$ and pseudoscalar nonet fields
$\pi_a$ are all included. $g_{ch}$ is the coupling constant of
chiral fields, its value is determined by the relation
\begin{equation}
\frac{g_{ch}^2}{4\pi}=\frac{9}{25}
\frac{m_u^2}{M_N^2}\frac{g_{NN\pi}^2}{4\pi},
\end{equation}
and $g_{NN\pi}$ is taken to be the experimental value.

In this model, we also employ an effective OGE interaction to govern
the short range behavior and a confinement potential to provide the
nonperturbative QCD effect in the long distance. Therefore the total
Hamiltonian of the system can be written as follows:
\begin{equation}
H=\sum_{i}T_{i}-T_{G}+\sum_{i<j}\left[V_{qq}(ij)+V_{q
\bar{q}}(ij)\right].
\end{equation}
The interaction between two quarks is expressed as follows:
\begin{equation}
V_{qq}(ij)= V^{OGE}_{ij} + V^{conf}_{ij} + V^{ch}_{ij},
\end{equation}
and the interaction between quark and antiquark has two parts:
direct interaction and annihilation parts,
\begin{equation}
V_{q \bar{q}}(ij)=V^{dir}_{q \bar q}+V^{ann}_{q\bar q},
\end{equation}
with
\begin{equation}
V_{q\bar q}^{dir}=V_{q\bar q}^{conf}+V_{q\bar q}^{OGE}+V_{q\bar
q}^{ch},
\end{equation}
and
\begin{equation}
V_{q\bar{q}}^{ch}=\sum_{j}(-1)^{G_j}V_{qq}^{ch,j}.
\end{equation}
Here $(-1)^{G_j}$ represents the G parity of the $j$th meson. The
confinement potential is taken as the quadratic form. The chiral
field coupling interaction includes two parts: scalar part and
pseudoscalar part. Their expressions and corresponding parameters
can be found in Refs. \cite{huangf04prc1,huangf04prc2}.

The annihilation part is complicated. The $NK$ system is the
simplest one, because $u(d)\bar{s}$ can annihilate neither to gluons
nor to vacuum. From the $NK$ scattering calculation where the
annihilation of $u(d)\bar s$ into a kaon meson has been considered
\cite{huangf04prc1,huangf04prc2}, we found that the annihilation
interaction is unimportant in the low energy scattering processes.
For the system $NK\pi$, when we treat it as a molecule-like state,
the distance between two particles is comparatively long, thus the
relative momenta of $NK$, $N\pi$ and $K\pi$ would be in the low
energy region. In this sense, we can neglect the annihilation part
to get the approximate effective interaction of $NK$, $N\pi$ and
$K\pi$ by fitting their low energy scattering phase shifts.

\begin{figure}[htb]
\vglue 2.0cm \epsfig{file=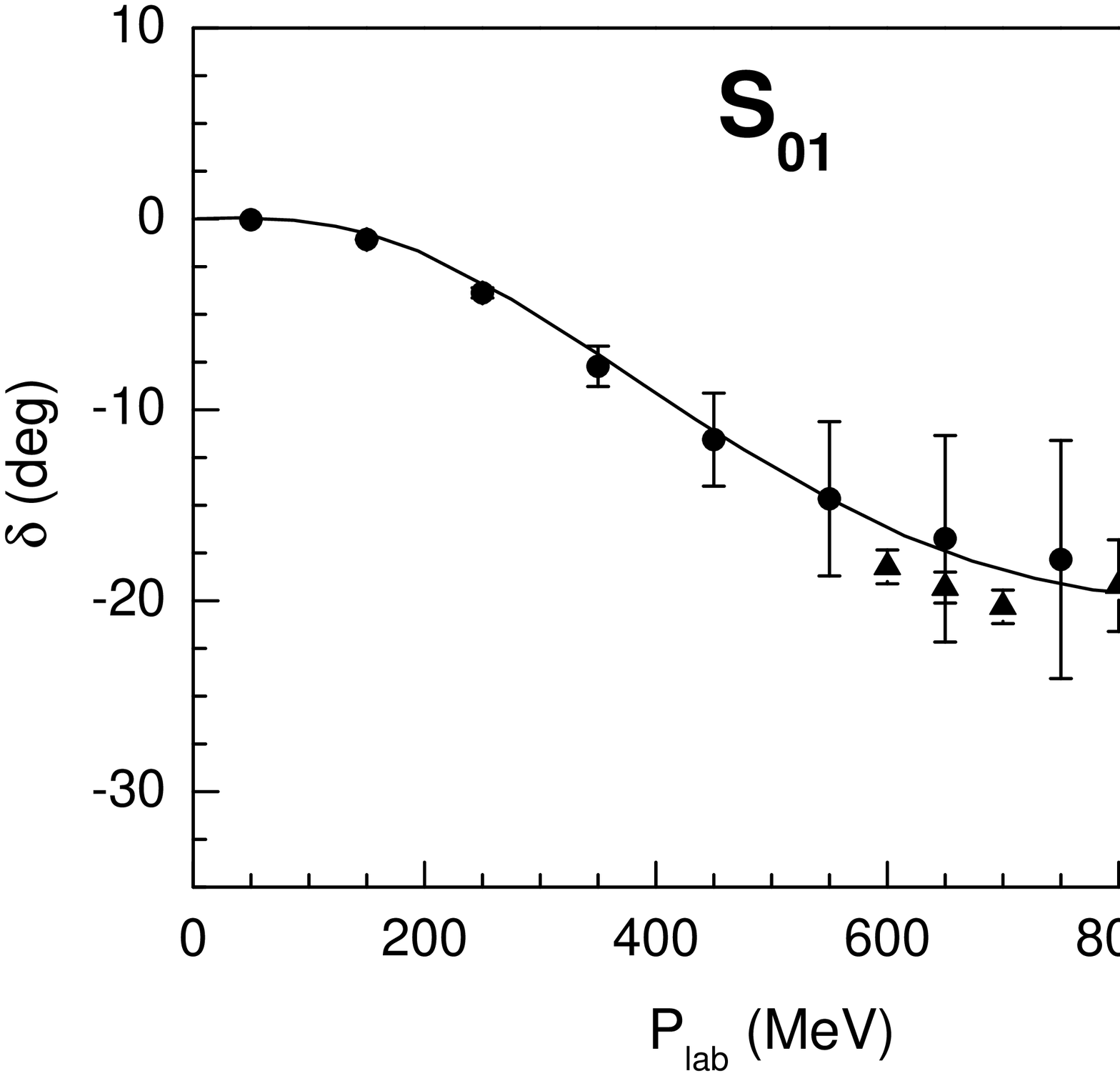,width=7.0cm}
\epsfig{file=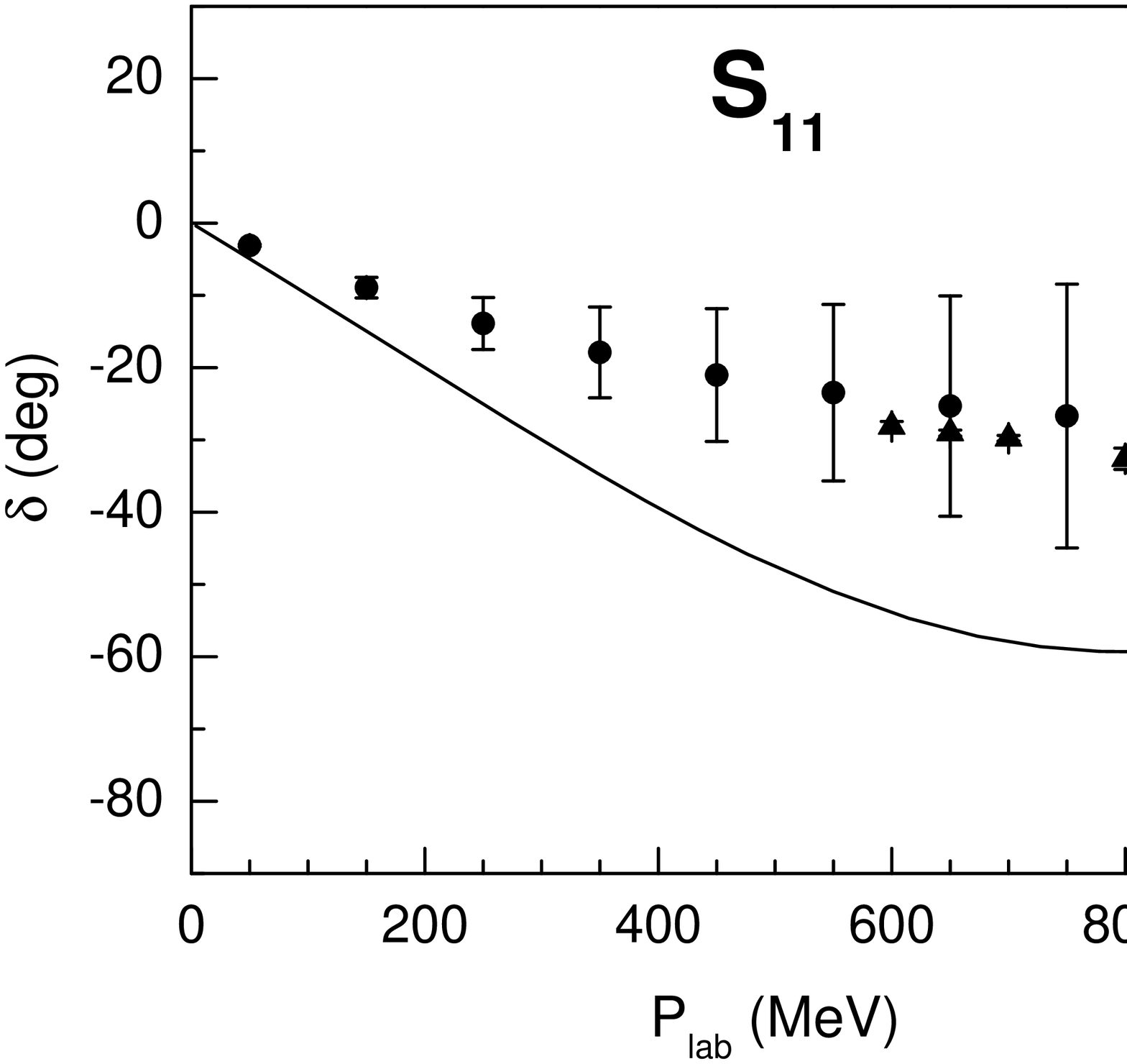,width=7.0cm} \vglue -2.5cm \caption{\small
\label{KN_phase} $KN$ $S$-wave phase shifts as a function of the
laboratory momentum of kaon meson. The first subscript denotes the
isospin quantum number and the second one twice of the total
angular momentum of the $KN$ system. The hole circles and the
triangles correspond respectively to the phase shifts analysis of
Hyslop {\it et al.} \cite{hyslop92} and Hashimoto
\cite{hashimoto84}.}
\end{figure}

\begin{figure}[htb]
\vglue 2.0cm \epsfig{file=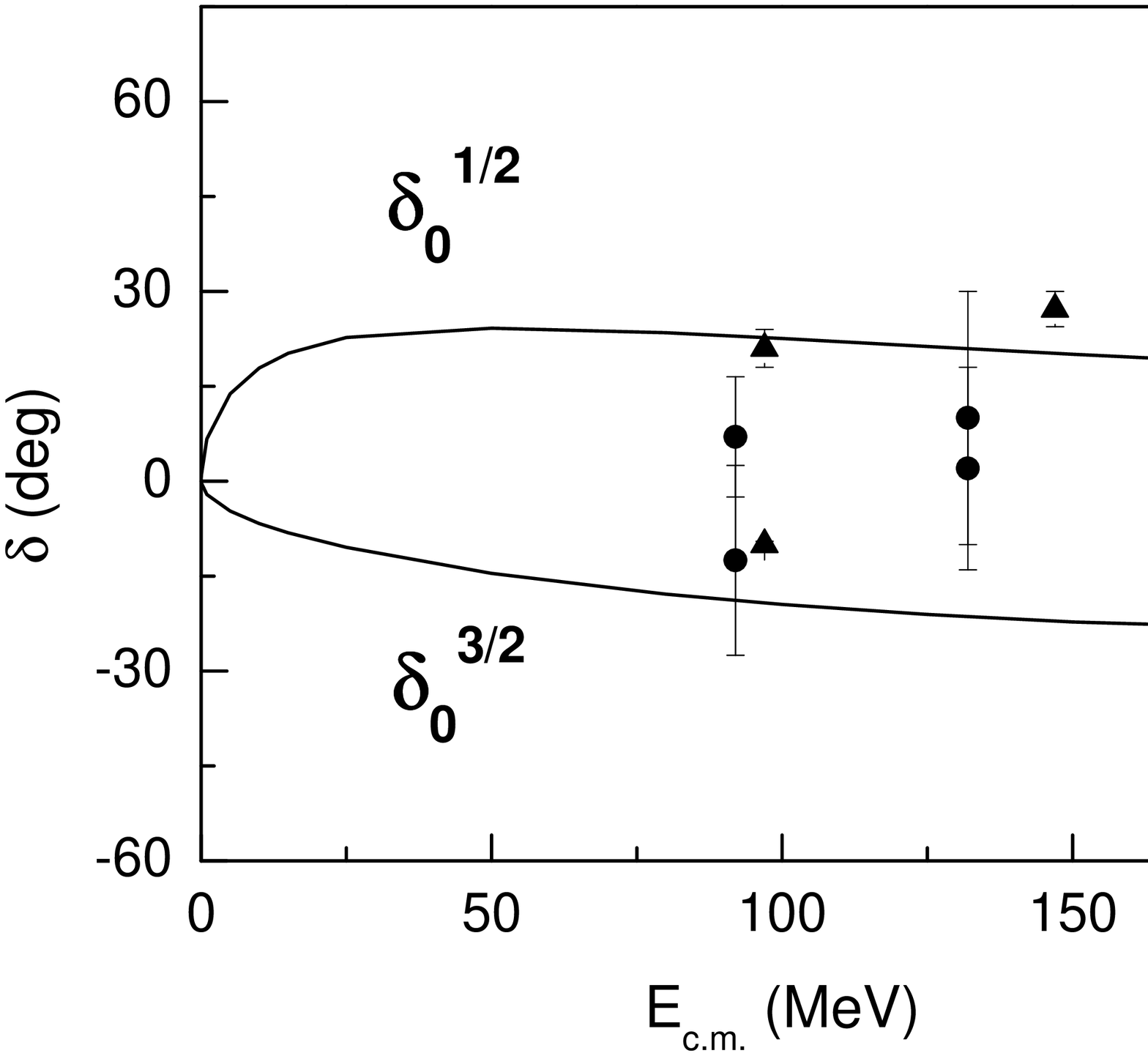,width=7.0cm}
\epsfig{file=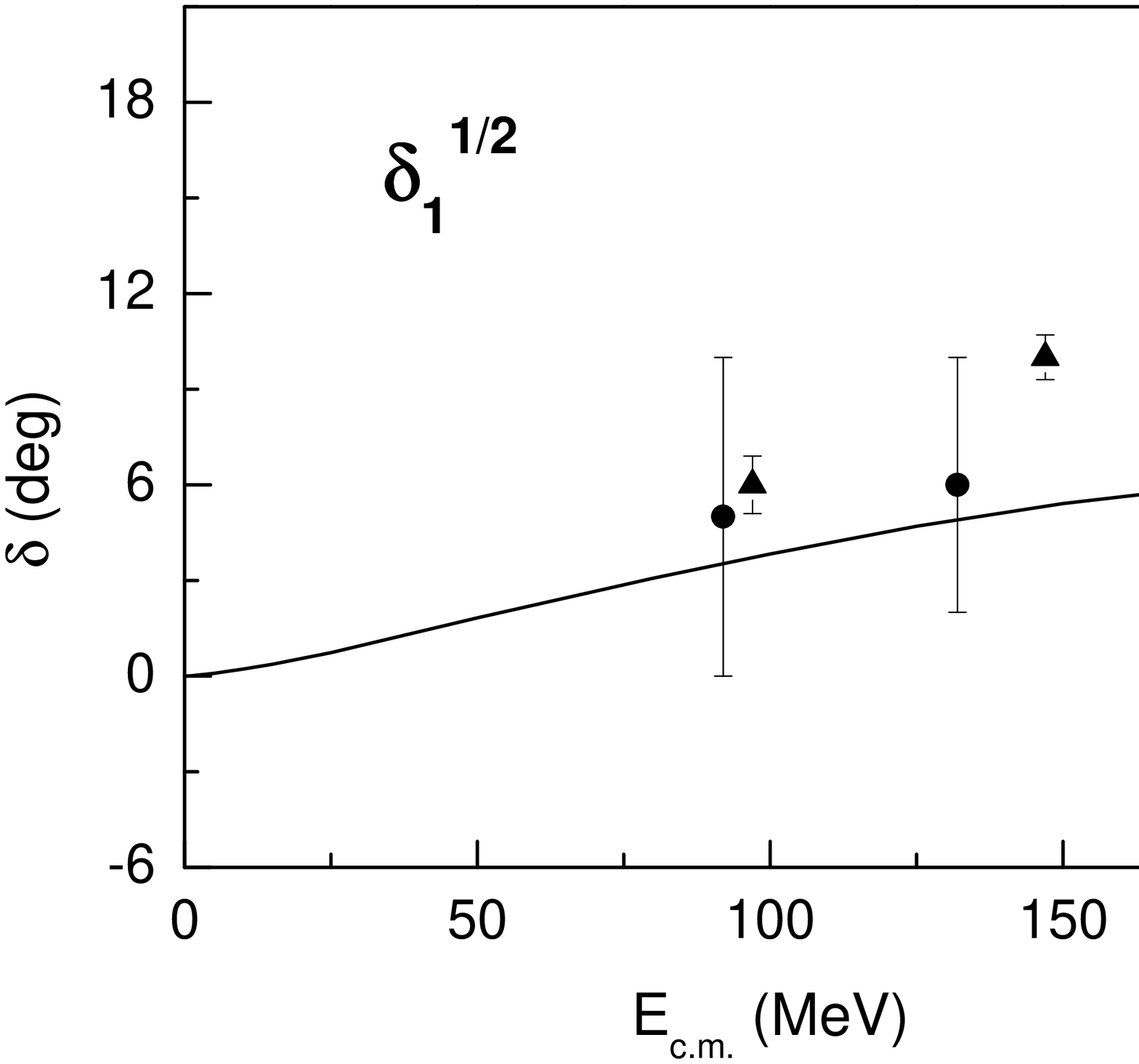,width=7.0cm} \vglue -2.5cm \caption{\small
\label{KPi_phase} $K\pi$ phase shifts as a function of the energy
of center of mass motion. The subscript denotes the orbit angular
momentum of the $K\pi$ relative motion and the superscript the
isospin of the $K\pi$ system. The hole circles and the triangles
correspond respectively to the phase shifts analysis of Mercer
{\it et al.} \cite{mercer71} and Estabrooks {\it et al.}
\cite{estabrooks78}.}
\end{figure}

\begin{figure}[htb]
\vglue 2.0cm \epsfig{file=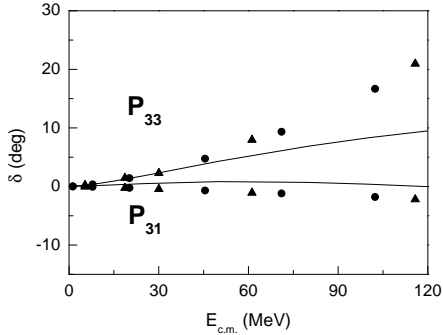,width=7.0cm} \vglue -2.5cm
\caption{\small \label{NPi_phase} $N\pi$ $P-$wave phase shifts as
a function of the energy of center of mass motion. The first
subscript denotes twice of the isospin quantum number and the
second one twice of the total angular momentum of the $N\pi$
system. The hole circles and the triangles correspond respectively
to the phase shifts analysis of Arndt {\it et al.} \cite{arndt95}
and Roper {\it et al.} \cite{roper65}.}
\end{figure}

Figs. 1-3 show the phase shifts of $KN$, $K\pi$, and $N\pi$. One
can see that in the low energy region the calculated phase shifts
are almost consistent with the experimental data.

Although the interactions of $KN$ $S$ wave are repulsive in both
isospin $I=0$ and $I=1$ channels, we notice that in the system
$(NK\pi)_{J^{\pi}=\frac{3}{2}^-,I=1}$, there are 1 pair of $N\pi$ in
the $P_{33}$ channel, $0.60$ pairs of $K\pi$ in $P$ wave with $I=
\frac{1}{2}$, and $0.28$ pairs of $K\pi$ in $S$ wave with $I=
\frac{1}{2}$. The calculated phase shifts indicate that all these
interactions are attractive. This shows that the $\pi$ plays a very
important role to glue the $K$ and the $N$ together, and as a
consequence, the energy of the state
$(NK\pi)_{J^{\pi}=\frac{3}{2}^-,I=1}$ can be very close to its
threshold.

We perform a rough estimate of the energy of the
$(NK\pi)_{J^{\pi}=\frac{3}{2}^-,I=1}$ system. The radial part of
the wave function is taken to be the harmonic oscillator form. In
this framework, the kinetic energy of the system is $2\omega$,
where $\omega$ is the harmonic oscillator frequency of the system.
In our calculation, we treat it as a parameter. Substituting
$M_N$, $M_K$, and $M_\pi$ by their experimental values in Eqs.
(1)-(4), we can obtain that in the
$(NK\pi)_{J^{\pi}=\frac{3}{2}^-,I=1}$ system there are:
\begin{eqnarray}
&&1~~~~~~~~\mbox{pair of}~~~~~(N\pi)_{l=1,j=\frac{3}{2},t=\frac{3}{2}},~~~~~\mbox{attraction}\nonumber\\
&&0.28~~~~\mbox{pairs of}~~~~(K\pi)_{l=0,j=0, t=\frac{1}{2}},~~~~~~\mbox{attraction}\nonumber\\
&&0.60~~~~\mbox{pairs of}~~~~(K\pi)_{l=1,j=1, t=\frac{1}{2}},~~~~~~\mbox{attraction}\nonumber\\
&&0.64~~~~\mbox{pairs of}~~~~(NK)_{l=0,j=\frac{1}{2},t=0},~~~~~\mbox{repulsion}\nonumber\\
&&0.32~~~~\mbox{pairs
of}~~~~(NK)_{l=0,j=\frac{1}{2},t=1},~~~~~\mbox{repulsion}
\end{eqnarray}
Where it can be seen from the phase shifts whether the interaction
is attractive or repulsive. When the frequency of the harmonic
oscillator, $\omega$, is taken to be around several tens MeV, the
calculated energy of $(NK\pi)_{J^{\pi}=\frac{3}{2}^-,I=1}$ is about
$70-80$ MeV higher than the threshold of $NK\pi$. We emphasize again
that this is just a very rough estimation, because the harmonic
oscillator wave function cannot offer an exact description of this
three-body system. And as is well known that the energy of the
system obtained from more accurate solution must be lower than that
from the estimation for a same Hamiltonian.

Because for the $(N\pi)_{jt=\frac{3}{2}\frac{3}{2}}$ system the
effect from the $\Delta$ resonance state is very important, we would
now like to consider the coupling of the configuration $(\Delta
K)_{J^{\pi}=\frac{3}{2}^-,I=1}$ to the system
$(NK\pi)_{J^{\pi}=\frac{3}{2}^-,I=1}$. First, we notice that the
$\Delta K$ interaction is attractive in the isospin one channel
\cite{sarkar05,huangf04prc1}, and it's energy can be taken to be the
value near the $\Delta K$ threshold \cite{huangf04prc1}. Then we
calculate the matrix element of the $\Delta$ and $N\pi$ interacting
vertex
\begin{equation}
\sqrt{4\pi}\frac{f_{\Delta N\pi}}{m_{\pi}}(\vec{\sigma}_{\Delta
N}\cdot \vec{q})(\vec{\tau}_{\Delta N} \cdot \vec{\phi}).
\end{equation}
For simplicity, we also adopt the harmonic oscillator wave
function in the calculation. When the size parameter $b$ is chosen
to be around $0.8-1.0$ fm and the coupling constant $f^{2}_{\Delta
N\pi}$ is taken to be $0.29$, we get
\begin{equation}
<\Delta\mid\sqrt{4\pi}\frac{f_{\Delta
N\pi}}{m_{\pi}}(\vec{\sigma}_{\Delta N}\cdot
\vec{q})(\vec{\tau}_{\Delta N} \cdot
\vec{\phi})\mid(N\pi)_{IJ=\frac{3}{2}\frac{3}{2}}> \approx 30-50~
\mbox{MeV}.
\end{equation}
As a consequence, the energy of the $NK\pi$ system will be $10-20$
MeV reduced by the mixing with the configuration $(\Delta
K)_{J^{\pi}=\frac{3}{2}^-,I=1}$.

Here we mention that in the case of $(K\pi)_{jt=0\frac{1}{2}}$, the
effect of the $\kappa$ resonance should also be considered. Because
the mass of $\kappa$ is about 345 MeV higher than $m_{K}+m_{\pi}$,
which is much larger than the mass difference between $N\pi$ and
$\Delta$, thus the effect of the $\kappa$ resonance must be smaller
than that of the $\Delta$ resonance. As an approximation, we neglect
it in this work.

In this article, we have estimated the energy of the $NK\pi$ system
with $J^{\pi}=\frac{3}{2}^-$ and $I=1$. By fitting the corresponding
phase shifts in the low energy region, the $KN$, $N\pi$ and $K\pi$
interactions are obtained from the chiral SU(3) quark model. The
estimated energy of $(NK \pi)_{J^{\pi}=\frac{3}{2}^-,I=1}$ is about
$70-80$ MeV higher than the threshold of the $NK\pi$ system. After
considering the mixing with the configuration $(\Delta
K)_{J^{\pi}=\frac{3}{2}^-,I=1}$, the energy of this system can go
down about $10-20$ MeV. The present calculations are rough and yet
do not produce a bound state of the three body system. However,
further refinements in the interaction and the method of calculation
could lead to such a state and it is worth calling the attention to
such a possibility. In fact recent, also qualitative, calculations
with a different quark model also suggest that the $\Theta^+$ could
be a $\frac{3}{2}^-$ state \cite{nam05}.

As mentioned, a new resonance $\Theta^+$ with $M_{\Theta}=1540$ MeV
and $\Gamma_{\Theta}<25$ MeV has been observed by several labs since
2003 \cite{hicks05}. Although there are also some experimental
groups who have reported negative results \cite{hicks05,woods05},
research is ongoing. This is because the $\Theta^+$ has strangeness
quantum number ${\cal{S}}=+1$, so that if it does exist, it must be
at least a 5-quark system, and if it can be explained as a
pentaquark it will be the first multi-quark state people found. At
the same time there are many theoretical works trying to explain the
structures and the properties of the $\Theta$ particle with various
quark models or other approaches \cite{zhusl04}. Because the mass of
$\Theta$, $M_{\Theta}$, is higher than the threshold of nucleon-kaon
system, $ M_N + M_K $, it is not easy to understand why its width is
so narrow, unless it has very special quantum numbers. As to the
mass of $\Theta$, although it is predicted by the original chiral
soliton model \cite{diakonov97} quite well, it is difficult to
generate the experimental $\Theta$'s mass and understand its narrow
width by the reasonably dynamical calculation based on the
constituent quark model when it is treated as a pentaquark state
\cite{stancu03,huangf04plb}. In Ref. \cite{huangf04plb}, we
calculated the energies of 17 lowest five-quark configurations in
the framework of the chiral quark model, and found that for both
cases of $J^{\pi}=\frac{1}{2}^-$ and $J^{\pi}=\frac{1}{2}^+$, their
energies are always about $250-350$ MeV higher than the experimental
value. It seems that either the $\Theta$ does not exist, which is
consistent with the new high statistical experiments \cite{woods05},
or the structure of this particle has to be understood in some other
mechanisms. If the report from RIHC-STAR group is true, i.e., there
is a $\Theta^{++}$ with ${\cal{S}}=+1$ and $M_{\Theta}\simeq 1530$
MeV \cite{rhic05}, we suppose that this $\Theta^{++}$ particle might
be explained as a three-body molecule-like state of $NK\pi$ with
$J^{\pi}=\frac{3}{2}^-$ and $I=1$. This is because: (1) its energy
is close to the threshold of $NK\pi$, thus it can be expected to be
a bound state; (2) its decay width must be quite narrow, because
when and only when the pion of the system is absorbed by the
nucleon, it can decay to $K$ and $N$; and (3) it is not easy to be
formed in the $K+N$ process. Though the above opinion is only a
qualitative discussion, we think more accurate study of the
structure and the properties of this system is worth doing in the
future work.

\vspace{0.5cm}

This work was supported in part by the National Natural Science
Foundation of China, Grant No. 10475087.

\end{document}